# How a Simple Chloroplast *psbA* Gene Mutation Changed World Agriculture

Jack Dekker
New Weed Biology Laboratory, Appleton, WI, 54911, USA
Email: newweedbiolab@gmail.com

## Contents

## Abstract

Atrazine as a weed control tactic profoundly changed world agriculture. Long-term use of this selective herbicide revealed resistant biotypes (R) with a single base pair mutation of the chloroplast *psbA* gene. The R phenotype emerged from a sequential cascade of pleiotropic effects from the plastid thylakoids to the whole plant. This pleiotropic reorganization of the R biotype revealed how photosynthesis was regulated at different levels of plant organization. The environment affected R plant productivity differently than in the susceptible (S) biotype. A consistent, differential, pattern of photosynthesis was observed between R and S over the diurnal light period. Photosynthetic superiority of a biotype was a function of the time of day, the age of the plant and the temperature of the environment.  Under highly favorable environmental conditions S often had the advantage over R.  Under certain less favorable conditions, stressful conditions, R can be at an advantage over S.  Pleiotropic reorganization revealed a sun-air-leaf (Shannon) communication system, providing insights into the complex interaction of chloroplast components in photosynthetic regulation.  Altered plastid thylakoid and stomatal function regulate how the R leaf utilizes the sun-air environment.  Movement of sun-air messages through the environment-plant channel demonstrated how raw sunlight and air are modified to a usable message for carbon fixation in signal transduction and utilization.  These insights showed how agriculture changed weed populations, and then how resistant weed populations changed human agriculture.  The changes in agriculture profoundly changed herbicide resistance: the development and introduction of commercial herbicide resistant crops (HRC).  The development of herbicide resistant crop cultivars extended the evolutionary reach of resistant weeds.  Previously unknown R weed biotypes were naturally selected in these introduced HRC's: an evolutionary spiral of human technology extended the phenotypic reach of R biotypes.  Introduction of the herbicide atrazine as a weed control tactic profoundly changed world agriculture in many anticipated and unanticipated ways: it was the advent of an unrecognized pandemic.

## 1.0  s-Triazine resistant plants: An agricultural pandemic

Introduction of the herbicide atrazine as a weed control tactic profoundly changed world agriculture in many anticipated and unanticipated ways: it was the advent of an unrecognized pandemic.  Resistant biotypes (R) of several important weed species were uncovered with the long-term use of atrazine as a primary weed control tactic in major crops like maize.  Elucidation of the mechanisms underlying atrazine resistance revealed a profound reorganization of the plant (Dekker, 2016b).

With the mutation of the chloroplast *psbA* gene a sequential cascade of pleiotropic effects led to the resistant biotype.  Elucidation of these changes revealed how photosynthesis was differentially regulated at many different levels of plant organization: from plastid thylakoid to leaf.  Elucidation of the pleiotropic changes revealed how the environment affected R and susceptible (S) plant productivity: the sun-air-leaf (Shannon) communication system.  Altered plastid thylakoid and stomatal function regulate how the R leaf utilizes the sun-air environment leading to carbon fixation, plant growth and development.

Control of how the R sun-air message is utilized allowed insight into how agriculture changed weed populations, and then how resistant weed populations changed human agriculture.  The changes in agriculture stimulated by the discovery of herbicide resistant weeds profoundly changed herbicide resistance: the development and introduction of commercial herbicide resistant crops (HRC).  The development of herbicide resistant crop cultivars extended the evolutionary reach of resistant weeds.  Previously unknown R weed biotypes were naturally selected in these introduced HRC's: an evolutionary spiral of human technology extended the phenotypic reach of R biotypes.

## 2.0  Evolutionary ecology of s-triazine resistant plants: Pleiotropic photosynthetic reorganization in the chloroplast *psbA* gene chronomutant

The pleiotropic reorganization of photosynthesis in R provided insights into the  evolutionary ecology of s-triazine resistant plants.  The reorganized R plant interacts with the environment in a different way than the susceptible biotype. Under environmental conditions highly favorable to plant growth, S often has the advantage over R.  Under certain less favorable conditions plant growth, stressful conditions, R can be at an advantage over S.  Control of this differential regulation of photosynthesis occurs at several different levels of plant organization, from the mutant *psbA* gene to the leaf.

**Figure 1.** Cascade of pleiotropic effects at the plastid, cell, tissue, organ, plant, population and community levels of plant organization consequential to the mutation of the *psbA* chloroplast gene.

| Gene | *psbA* chloroplast gene codon 264 mutation (R) | | |
|---|---|---|---|
| | ↓ | | |
| Chloroplast /Cell | altered chloroplast D-1 protein structure: s-triazine herbicide resistance | | |
| | changes in PSII electron transport | changes in chloroplast membranes:<br>  -increased thylakoid membrane stacking<br>  -increased thylakoid grana lamellae fatty acid unsaturation<br>  -greater low temperature lipid fluidity | |
| | | ↓ | |
| | | tolerance to cool temperatures<br>tolerance to temperature change | |
| | ↓ | | |
| Leaf/Plant | R is a chronomutant | | |
| | R photosynthetic mutants | R stomatal mutants | |
| | ↓ | ↓ | |
| | differential R/S diurnal pattern of chlorophyll fluorescence | R increased stomatal aperture opening | R equal/greater than S:<br>-H$_2$O conductance<br>-intercellular CO$_2$ |
| | ↓ | ↓ | |
| | differential R/S diurnal pattern of photosynthetic carbon assimilation | cool TEMP leaf:<br>-R equal/cooler than S<br>-R/S carbon assimilation equal | high TEMP leaf:<br>-R leaves cooler<br>-S leaves warmer |
| | Younger plants | Older plants:<br>Age reversal with onset of plant reproduction: | ↓ |
| | R greater than S: early and late in the diurnal; older only late | R greater than S: early, midday; whole day | Stomatal function differentially regulates carbon assimilation via leaf cooling |
| | S greater than R: midday; whole day | S greater than R: only late in diurnal | ↓ |
| | ↓ | | |
| | R/S niche diversity | | |
| | R niche:<br>-young: early & late in diurnal<br>-older: early, midday; all day | S niche:<br>-young: midday, all day<br>-older: only late | shade-adapted leaf morphology |
| | ↓ | | |
| Population | enrichment of R biotypes | | |
| | selective advantage with s-triazine herbicide use | enhanced intra-specific photosynthetic phenotype-niche diversity | increased seed dormancy |

## 2.1 Gene-Chloroplast

The nature of s-triazine resistant plants (R) is a complex adaptive photosynthetic system arising as a consequence of a single base pair lesion in the chloroplast *psbA* gene (Dekker, J. 2016a, b). A single base-pair substitution at codon 264 in the *psbA* chloroplast gene in the highly conserved photosynthetic apparatus leads to a cascade of changes in the plants morphology, physiology and ecological reaction to its immediate environment, a non-intuitive result. With the genetic lesion, a dynamic re-organization of interacting, interdependent, functional plant units occurs. A new homeostatic equilibrium emerges among the plastids, cells and organs of the whole plant R phenotypes (figure 1). Order in this highly conserved photosynthetic system is emergent.

### 2.1.1 D-1 protein.
The D-1 protein product of *psbA* is a key element of photosystem II electron transport. Altered electron transport in R causes a pleiotropic cascade of self-reorganization of interacting, interdependent, functional traits. This adaptive reorganization of photosynthetic components in the chloroplast may be a compensatory mechanism to maintain a functional interaction of the PS II complex lipids and proteins (Pillai and St. John, 1981).

The pleiotropic cascade of functional and structural changes conferred by changes in the D-1 protein could imply that the amino acid substitution is close to a primary functional and structural source of photosynthetic regulation. Mattoo (Mattoo et al., 1984) has suggested that the rapid anabolism-catabolism rate of the D-1 protein could serve as a signal resulting in the reorganization of membranes around the PSII complex. This dynamic reorganization has consequences for evaluating and understanding regulatory effects of electron transport in carbon assimilation (Dekker, 1993).

### 2.1.2 Chloroplast lipids.
R also possesses tolerance to cool temperatures, and to changes in temperature, which is conferred by pleiotropic membrane lipid changes. At relatively cool temperatures it has been hypothesized that the change in lipid saturation of chloroplast membranes could confer cold tolerance to R plants, resulting in greater carbon assimilation rates in R under those conditions (Pillai and St. John, 1981).

It is hypothesized that these ontogenetic and diurnal patterns of differential photosynthesis may be a consequence of correlative diurnal fluctuations in fatty acid biosynthesis and the dynamic changes in membrane lipids over the course of the light-dark daily cycle, changes in leaf membrane lipids with age (Lemoine et al., 1986), or microenvironmental temperature influences (Ireland et al., 1988; Dekker and Burmester, 1990; Ducruet and Lemoine, 1985).

## 2.2 Leaf-Plant

### 2.2.1 Chronomutant.
A consistent, differential, pattern of many photosynthetic functions was observed between R and S *Brassica napus* over the course of a diurnal light period. Photosynthetic superiority of one biotype to another was a function of the time of day, the age of the plant and the temperature of the environment. Many plant species exhibit an endogenous rhythm of carbon assimilation and stomatal function once entrained in a photoperiod. This rhythm is regulated to some extent independently of the plant's direct response to PPFD (Browse et al., 1981).

### 2.2.2 Diurnal chlorophyll fluorescence.
Our studies revealed a consistent, differential, pattern of Chl *a* fluorescence ($F_l$) (Dekker and Westfall, 1987a, b). A phase shift in leaf chlorophyll fluorescence (LCF) maxima occurred in the daily light-dark cycle, supporting the hypothesis that alterations in chloroplast structure conferring s-triazine resistance imply altered temporal behavior of photosynthetic activity.

### 2.2.3 Diurnal carbon assimilation.
The nature of R plants is how local photosynthetic opportunity spacetime is seized and exploited relative to that of the susceptible (S) wild type. The altered D-1 protein product of the *psbA* gene has been regarded as less photosynthetically efficient in the R biotypes of several species. Past studies have shown lower carbon assimilation (A) and whole plant yields in R relative to S; in other comparisons R was greater than S; in still others R and S were comparable (refs).

Studies were conducted observing diurnal patterns of A during development. Our studies revealed a consistent, differential, pattern of carbon assimilation between R and S. Younger R plants had greater photosynthetic rates early and late in the diurnal, while those of S were greater during midday as well as the photoperiod as a whole. As *B. napus* plants aged, differences in A between biotypes increased: only during the late diurnal period was R greater.

As *B. napus* began reproductive development, a reversal of photosynthetic differences occurred. R carbon assimilation was greater than S early, midday, and for the whole day period; S was superior to R only late in the day.

These results indicate a more complex pattern of photosynthetic carbon assimilation than previously reported. The photosynthetic superiority of a biotype is a function of the time of day and the age of the plant.

**2.2.4 Stomatal function.** R plants differ in stomatal function from S plants. R plants have markedly different leaf and stomatal responses to temperature. Our studies revealed a consistent, differential, pattern of leaf temperature, total conductance to water vapor (g), and leaf intercelluar $CO_2$ partial pressure ($C_i$) (Dekker, 1993; Dekker and Burmester, 1992; Dekker and Westfall, 1987 a, b) between S and R *Brassica napus* over the course of a diurnal light period: R is a chronomutant. Total conductance to water vapor and intercellular $CO_2$ partial pressure in R was either equal to, or greater than, S over the lifetime of those plants, with the possible exception of some atypical episodes late in ontogeny and late in the light period. In addition to other pleiotropic effects, R plants appear to be stomatal mutants. They constitute a model system to study regulation of stomatal function and the relationship between environmental cues and stomatal behavior.

**2.2.5 Leaf temperature.** As a consequence of these phenomena, leaves of R plants were either the same temperature, or cooler, than leaves of S plants for the entire lives of both biotypes. At lower leaf and air temperatures R and S leaves function in a similar way. As air temperature increases, their responses diverge, allowing R plants compensate for their sensitivity to high temperature. R leaves generally are cooler and total conductance to water is greater than in S, probably due to greater stomatal aperture size. As a result, at higher air temperatures R leaves photosynthesize at cooler leaf temperatures closer to the optimal for both biotypes.

**2.2.6 Carbon assimilation.** Stomatal function differentially regulates carbon assimilation in these two biotypes. R is heat tolerant: R high leaf temperature sensitivity is compensated by higher stomatal conductance leaf cooling. R possesses greater heat tolerance than S due to leaf cooling from greater stomatal conductance and leaf intercellular $CO_2$ partial pressure. As a consequence, at all important physiological temperatures (10-35° C) R leaves are cooler than S leaves.

R exhibited high temperature sensitivity: when high leaf temperature (e.g. 35° C was closely controlled carbon assimilation was much lower than S; consistent with others (Ducuret and Ort, 1988; Havaux, 1989; Gounaris and Barber, 1983). When leaf temperature was not directly controlled, but air temperature was, R carbon assimilation exceeded that of S at relatively high temperatures (e.g. 35° C air temperature) (Dekker and Burmester, 1990, 1992). In both experimental conditions R leaf stomatal conductances were usually greater than in S.

**2.2.7 Shade adaptation.** Pleiotropic changes in R result in chloroplast and whole leaf morphology similar to that of low light, dark-adapted, plants. The acquisition of shade-adapted morphology in R is not a plastic response of the phenotype to environment, but a constitutive consequence of the pleiotropic cascade. Shade tolerance is a complex, multi-faceted property of plants. Different plant species exhibit different adaptations to shade, and a particular plant can exhibit varying degrees of shade tolerance, or even of requirement for light, depending on its history or stage of development.

## 2.3 Photosynthetic regulation

Studies were conducted to determine the response of R and S to different temperatures and gas atmospheres with infrared gas analysis and pulse amplitude modulated chlorophyll fluorescence techniques. Photosynthetic regulation can be separated into three categories based on these studies (Dekker, et al., 1990; Dekker and Sharky, 1990, 1992).

The first category is Rubisco-limited photosynthesis. Studies were conducted to test if there is any regulatory role for ribulose-1,5-bisphosphate carboxylase/oxygenase (Rubisco) in photosynthetic carbon assimilation in R and S *Brassica napus*. When carbon assimilation was Rubisco-limited there was little difference between R and S biotypes. Rubisco percent activation and initial activity may account for R and S carbon assimilation (A) differences during midday. Differences in A early and late in the day were not accounted for by differences Rubisco (initial, total, % activation).

A second category, feedback-limited photosynthesis, was most evident at 15°C, less so at 25°C.

The third category, photosynthetic electron transport-limited photosynthesis, was evident at 25 and 35° C. R exhibited much more electron transport-limited carbon assimilation at 25°C and 35°C than did S. At 15°C neither R nor S exhibited electron transport limited carbon assimilation. The primary limitation to photosynthesis changes with changes in leaf temperature: electron transport limitations in R may be significant only at higher temperatures.

### 2.3.1 R adaptation to the environment and regulation of carbon assimilation.
Regulation of photosynthesis in R and S are controlled by many different factors. Limitations in electron transport in R is not the only critical factor in yield losses at the whole plant level. The pleiotropic effects observed in R result in a new equilibrium between functional and structural components. It is this new dynamic pleiotropic reorganization that regulates carbon assimilation in R.

Electron transport limitations are only one possible regulatory point in the photosynthetic pathway leading from light-harvesting and the photolysis of water, through ribulose bisphosphate carboxylase/oxygenase, to starch/sucrose biosynthesis, translocation, and utilization. Carbon flux through the leaf is regulated at many points. Electron transport, even in R, is not the only critical regulatory step. In fact, Dekker and Sharkey (1992) have shown that the primary limitation to photosynthesis changes with changes in leaf temperature, and that electron transport limitations in R may be significant only at higher temperatures.

### 2.3.2 Photosynthetic niches.
It can be envisioned that there were environmental conditions prior to the introduction of s-triazine herbicides in which R had an adaptive advantage over the more numerous S individuals in a population of a species: in cool, low-light environments early and late in the day; at high temperatures; and late in the plant's development. The reorganized R plant interacts with the environment in a different manner than does S. Under environmental conditions highly favorable to plant growth, S often has an advantage over R. Under certain less favorable conditions to plant growth, stressful conditions, R can be at an advantage over S.

Under certain conditions R might have exploited a photosynthetic niche under-utilized by S. These conditions may have occurred in less favorable environments and may have been cool (or hot), low light conditions interacting with other biochemical and diurnal plant factors early and late in the photoperiod, as well as more complex physiological conditions late in the plant's development.

Under these conditions R survival and continuity could have been ensured at a higher frequency of occurrence than that due to the mutation rate of the *psbA* plastid gene alone, independent of the existence of a speculated plastome mutator (Duesing and Yue, 1983).

**2.3.3 Sun-air-leaf signal transduction.** The dynamic changes in photosynthetic functions in R and S may be better understood by analysis of the *psbA* chronomutant sun-air-leaf (Shannon) communication system. Movement of energetic equivalents through the environment-plant channel reveal how components of sunlight and air are absorbed, transmitted and utilized as usable messages for carbon fixation through each of the altered R pleiotropic loci.

## 3.0 The altered sun-air-leaf communication system in s-triazine resistant chronomutants

The communication of sun-air substrates into the R plant leaf provides insights into the evolutionary ecology of s-triazine resistant plants. This Shannon communication system (Shannon and Weaver, 1949) was revealed with the elucidation of pleiotropic reorganization of photosynthesis in R chronomutants. The sun-air-leaf communication system provides insights into the complex interaction of chloroplast components in photosynthetic regulation; from sun-air substrates to gene-leaf growth.

> "What lies at the heart of every living thing is not a fire, not a warm breath, not a 'spark of life'. It is information, words, instructions ... If you want to understand life, don't think about vibrant, throbbing gels and oozes, think about information technology." (Dawkins, 1986).

**3.1 Biological information.** The nature of weeds is an environment-biology communication system. Biology is information. Evolution itself embodies an ongoing exchange of information between organism and environment. For biology, information comes via evolution; what evolves is information in all its forms and transforms. Information is physical. Biology is physical information with quantifiable (Kolmogorov) complexity. The gene is not the information-carrying molecule, the gene is the information.

Information in biological systems can be studied. An important problem in science is to discover another language of biology, the language of information in biological systems. "The information circle becomes the unit of life. It connotes a cosmic principle of organization and order, and it provides an exact measure of that (Loewenstein, 1999). This information circle for weedy plants is the predictable developmental events of the annual life history.

**3.2 Shannon communication systems.** Information theory was developed by Claude E. Shannon (Shannon and Weaver, 1949) to find fundamental limits on signal processing operations such as compressing data and on reliably storing and communicating data. Since its inception it has broadened to find applications in statistical inference, networks (e.g. evolution and function of molecular codes, model selection in ecology) and other forms of data analysis. The soil-weed seed communication system was first described for weedy *Setaria* sp. (Dekker, 2013). A Shannon communication system of any type (e.g. language, music, arts, human behavior, machine) must contain the following five elements (E) presented in figure 2, as well as message and signal.

A Shannon communication system includes these elements, as well as the concepts of message and signal. A message is the object of communication; a vessel which provides information. It can also be this information, its meaning is dependent upon the context in which it is used. A signal is a function that conveys information about the behavior or attributes of some phenomenon in communication systems. Any physical quantity exhibiting variation in time or variation in space is potentially a signal if it provides information from the source to the destination on the status of a physical system, or conveys a message between observers, among other possibilities.

**Figure 2.** Schematic diagram of Shannon information communication system (Shannon and Weaver, 1949).

|   | Communication Element | Description |
|---|---|---|
| 1 | Information source | entity, person or machine generating the message (characters, math function) |
| 2 | Transmitter | operates on the message in some way: it converts (encodes) the message to produce a suitable signal |
| 3 | Channel | the medium used to transmit the signal |
| 4 | Receiver | inverts the transmitter operation: decodes message, or reconstructs it from the signal |
| 5 | Destination | the person or thing at the other end |
|   | Concepts |   |
| Message | | the object of communication; either providing information, or it is the information |
| Signal | | a physical function that conveys information about the behavior or attributes of some communication system phenomenon; a physical quantity exhibiting variation in space-time |

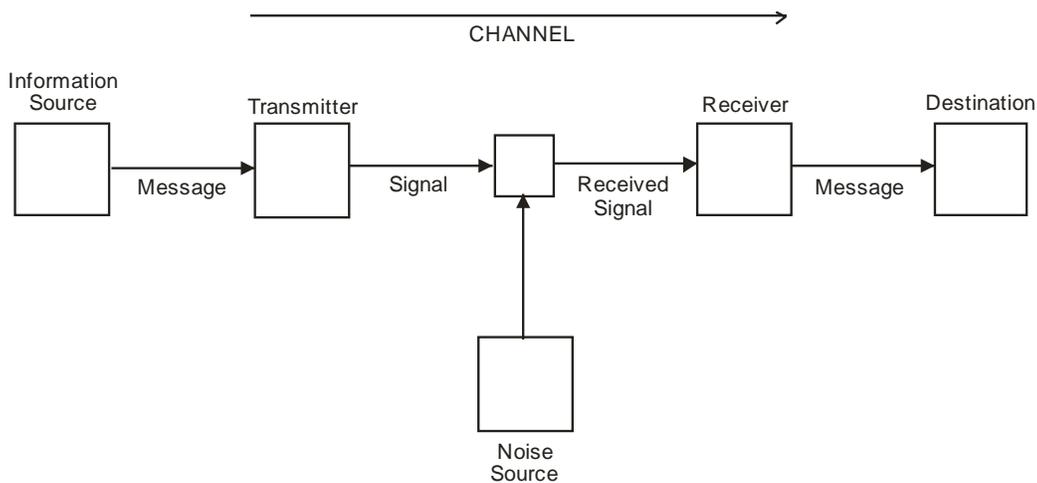

**3.3 The sun-air-leaf communication system.** The nature of weedy s-triazine resistant plant life history is a complex adaptive, photosynthetic communication system arising from its component functional traits (figure 3). Functional traits controlling leaf photosynthetic behavior are physical information that has evolved in ongoing communication between organism and environment leading to local adaption.

The sun-leaf communication system provides insights into the complex interactions of the environment and the leaf-chloroplast components in photosynthetic regulation (figures 3, 4).

Communication of the sunlight-air message through s-triazine resistant and susceptible (R/S) plants ends when fixed carbon becomes available for growth and development. At each point in this sequential communication system photosynthetic regulation can occur.

**3.3.1 Communication channel (E3): atmospheric sunlight into the leaf.** The medium used to transmit the signal in the R/S sun-leaf communication system originates with solar radiation passing through the atmosphere to the plant leaf. Impinging on the leaf the channel proceeds through the leaf stomatal openings to the interior tissues, cells and chloroplast. Photosynthetic activity resulting in fixed carbon (e.g. sucrose) occurs in the chloroplast thylakoids. The destination of the fixed carbon produced is plant growth and development.

**3.3.2 Information source (E1): atmospheric solar radiation.**

> "The *information source* selects a desired *message* out of a set of possible messages" (Shannon and Weaver, 1949)

**Figure 3.** The Shannon environmental-biological sun-air-leaf (photo-thermal-aero-hydro time) communication system between the sun and the plant leaf contains the five elements (E) and components (signal, message, noise) that interact sequentially through the communication channel (PPFD, photosynthetic photon flux density).

| Shannon Sun-Air-Leaf Communication System | | | | |
|---|---|---|---|---|
| Element-Message-Signal | | | | E3 Channel |
| E1 Information source | solar radiation, air | | | |
| Message | E1 → 2: sunlight, heat, $CO_2$, $H_2O$ | | | |
| E2 Transmitter | leaf stomata opening | | | |
| Transmitted Signal | E2 → noise: PPFD, $CO_2$, $H_2O$, heat near leaf | | | |
| Noise | Photo-thermal-aero-hydro time environment | | | |
| | Light Diurnal: early midday late | Heat: cool hot | Leaf Age: young old | 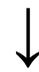 |
| Received Signal | Noise → E4: reduced photosynthetic equivalents (PSII) | | | |
| E4 Receiver | chloroplast thylakoid membranes: carbon fixation photolysis of water | | | |
| Received Message | E4 → E5: reduced, fixed carbon (sucrose) | | | |
| E5: Destination | plant growth and phenotype development | | | |

The source of information generating the message in the R/S sun-leaf communication system is the sun passing through the atmosphere. This information flows in a continuous diurnal pattern affected by the local environment: including seasonal time, climate, site latitude-longitude-elevation-slope-aspect, agricultural and other human disturbances. This information flows to the plant leaf, which responds to only a limited portion of the entire solar radiation-atmospheric gases information, the message controlling behavior.

**3.3.3 Message (E1 to E2): sunlight-heat-carbon dioxide-water.** The message for the sun-air-leaf communication system is a subset derived from the more extensive information source (sun, air). This more specific message sent to the leaf transmitter (E2) includes sunlight, heat and atmospheric gases ($CO_2$, $H_2O$) required for continued growth and development of the plant: photo-thermal-$CO_2$-hydro time.

**3.3.4 Transmitter (E2): leaf sun-air absorption-exchange.**

> "The *transmitter* changes this *message* into the *signal* which is actually sent over the *communication channel* from the transmitter to the *receiver*." (Shannon and Weaver, 1949)

The transmitter is the leaf surface-stomata that converts/encodes/changes the message to produce a suitable signal. The sun-air message is transmitted to the plant via the leaf surface (sunlight, heat) and the stomata (air). Photosynthetic active radiation (PAR, visible light wavelengths between 400 to 700 nanometers) and heat are absorbed by the chloroplast thylakoid membranes in the leaf interior. Modulation of the stomatal opening in the leaf allows available gases to be exchanged between the leaf exterior and interior: excretion of oxygen, absorption of $CO_2$ for photosynthetic carbon fixation. Stomatal opening differs between R/S, the first point of differential photosynthetic regulation between the biotypes.

**Figure 4.** The Shannon photo-thermal-aero-hydro environmental-biological sun-air-leaf communication system between the sun and the s-triazine resistant plant leaf.

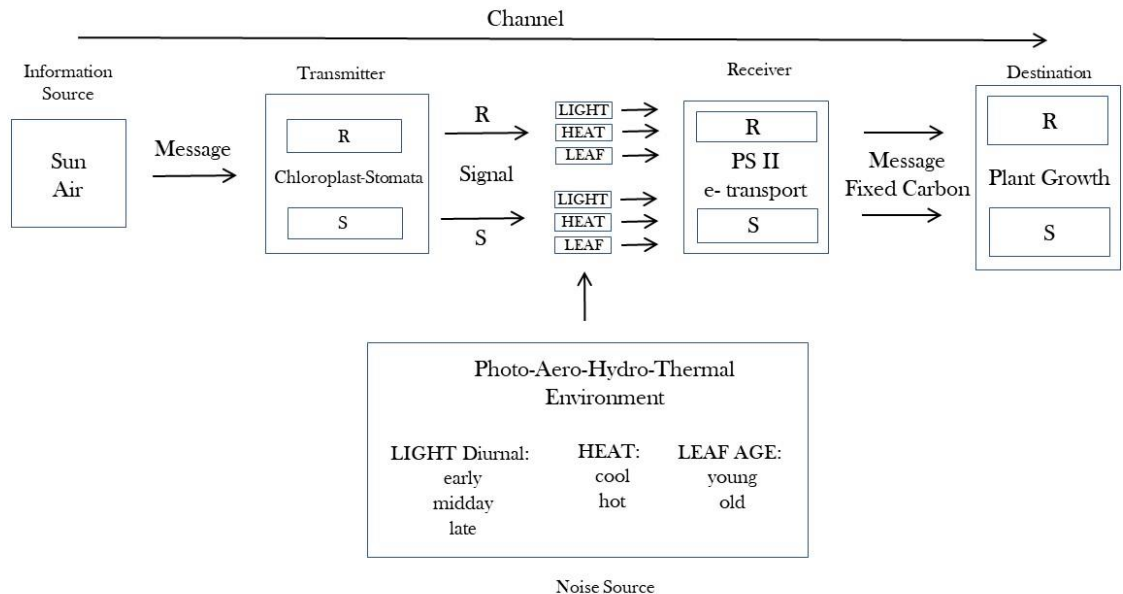

**3.3.5 Transmitted signal (E2 to noise): usable solar radiation-air.** Stomatal aperture opening allows the exchange of atmospheric gases between the leaf exterior and interior. Selective absorption of usable solar radiation by thylakoids is enabled by this exchange. The leaf transmitter converts solar radiation into usable PPFD (photosynthetic photon flux density; µmol quanta m$^{-1}$ period$^{-1}$) signal, the active spectrum of solar radiation which drives photosynthesis. Gaseous absorption into the leaf interior makes gaseous messages available as suitable signals for subsequent photosynthetic activities in the chloroplast (e.g. photolysis of water, carbon assimilation).

**3.3.6 Noise source: local photo-thermal-aero-hydro time environment.** Several sources of noise are introduced in the communication channel between the transmitter and receiver. Noise includes everything that corrupts the sun-air-leaf signal: physical interference, chemicals, localized areas of anomalous gas and water content/exchange; leaf health. Noise in the local photo-aero-hydro-thermal environment modulate and mitigate the transmitted signal. The transmitted signal is changed by these environmental sources of noise.

Introduction of noise in the communication channel means the received message exhibits increased uncertainty, divergence from the transmitted signal. This environmental modulation provides both photo-thermal-gas substrate for carbon fixation, as well as information allowing the leaf to optimize local resource limitations in the individual leaf, control of phenotype development.

Predictable signal noise components in the sun-air-leaf communication system include the light diurnal (time of day), heat and leaf age (figure 3). R/S biotypes differ in their responses to these predictable noise components. Noise uncertainty can result in inhibition (e.g. death, low humidity, heat, clouds, night, herbicides) or stimulation (e.g. moist air, moderate temperatures, clear sky, time of daylight). Experimentally the influence of these three predictable noise components between the transmitted and received signal has been quantified in studies reported. The received sun-air-leaf signal is selected from the more varied transmitted set.

**3.3.7 Received signal (noise to E4): reduced photosynthetic substrates.** The signal received from the transmitted signal is reduced photosynthetic equivalents providing energy to reduce carbon in the chloroplast. This received signal is generated by PSII electron transport and the photolysis of water.

**3.3.8 Receiver (E4): chloroplast thylakoid membranes.**

> "The *receiver* is a sort of inverse transmitter, changing the transmitted signal back into a message, and handling this message on to the destination." (Shannon and Weaver, 1949)

The receiver in the sun-air-leaf communication system is the chloroplast thylakoid membranes that fix carbon from energetic substrates provided by photosystem 2 (PS II) activity and the photolysis of water. The receiver reconstructs the transmitted signal, after noise modulation, into substrates usable by the plant for growth and development (e.g. sucrose).

**3.3.9 Received message (E4 to E5): reduced carbon.** The receiver generates the message sent to the plant destination: assimilated, fixed carbon.

**3.3.10 Destination (E5): plant growth.** Communication in the sun-air-leaf system is completed when the received message (fixed carbon) is utilized by the plant for growth and development. The communication process results in differential growth and development between s-triazine resistant and susceptible biotypes.

## 4.0 The extended phenotype in s-triazine resistant plants

Atrazine introduction profoundly changed world agriculture in many anticipated and unanticipated ways: an unrecognized pandemic. The changes in agriculture stimulated by the discovery of herbicide resistant weeds profoundly changed herbicide resistance: the development and introduction of commercial herbicide resistant crops. The evolutionary reach of resistant weeds extended the resistant phenotype with this pandemic: herbicide resistance changed agriculture, then agriculture changed herbicide resistance.

### 4.1 How herbicide resistance changed agriculture

Atrazine introduction profoundly changed world agriculture. Herbicides changed the phenotypic composition of weed populations: novel R biotypes appeared. In turn, resistant weed biotypes changed agriculture. New management strategies appeared, including commercialization of herbicide resistance. Changes in weed populations and weed management changed weed science.

**4.1.1 How weeds changed.** s-Triazine resistance was first discovered in late 1960's: one of the first documented instances of herbicide resistance. The early reports indicated R was photosynthetically inferior to S. Since that time R has been found throughout the world where atrazine is used in weed management. Atrazine is no longer a key herbicide in maize weed control systems. Since that time many other types of herbicide resistance have been documented in many weed species and cropping systems throughout the world.

**4.1.2 How agriculture changed.** s-Triazine resistant plants initiated a series of inevitable, consequential events that profoundly changed world agriculture. In the beginning herbicide use naturally selected resistant weed biotypes. Resistant weeds necessarily changed how weeds were managed, and created opportunity for commercialization of resistant traits in crop cultivars. Commercial opportunity led inevitably to the introduction and utilization of herbicide resistance (HR) in crop cultivars (Dekker and Duke, 1995; Dekker and Comstock, 1992). Crop-weed management systems based on HR technology centralized and simplified the herbicide-seed industry, notably glyphosate-resistant crops. Germplasm heterogeneity in several major world crops dramatically decreased with HR crop introduction. Subsequently resistant weed biotypes were naturally selected in the new introduced HR crops.

**4.1.3 How resistant weeds and crop cultivars changed agriculture.** Resistant weed discovery led to many changes in weed control tactics, and stimulated more sophisticated non-herbicidal weed management strategies. HR crop introduction led to a decrease in the numbers and types of herbicides used in those crops. The elimination of competitive herbicides led to a reduction in the use of many previously registered herbicides, and to the elimination and consolidation of companies developing new herbicides.

The first resistant crop cultivars developed were those tolerating s-triazine herbicides. Conventional breeding techniques were utilized to develop HR rapeseed (*Brassica napus*) in Ontario, Canada in the early 1980's (e.g. (Beversdorf and Hume, 1984; Beversdorf, Hume and Donnelly-Vanderloo, 1988; Dekker, 1983a-l). Little commercial development occurred with this type of crop cultivar resistance.

Transgenic crop development potential was realized in the early 1990's with the development of glyphosate resistant maize and soybeans. The economic impact of this technology profoundly changed maize and soybean management systems throughout the world. The natural selection of glyphosate resistant weeds inevitably followed the introduction of glyphosate R crops.

Social, anthropological, changes in the human agricultural value system emerged from the introduction of these new technologies. HR crop cultivar development now placed emphasis on "what's best for the farmer". The goal of the new commercialized herbicide-seed technologies was crop production efficiency. The agricultural goal of crop cultivar introduction was no longer "what's best for the consumer of crops". Crop cultivar quality and heterogeneity were of secondary importance. The result was the industrialization of crop production: ease and efficiency of production, and control of production via intellectual property rights to maximize profit.

**4.1.3 How weed science, weed evolutionary ecology, changed.** Herbicide resistance changed the way weed science is conducted, the questions formulated to provide answers to the emerging R pandemic. The emphasis in weed science shifted from weed control tactics to R weed management strategies.

Strategic management highlighted the need to embrace weed-crop ecology to address these new problems. The science of evolution has always been problematic in the USA, weed evolutionary biology has languished despite the role natural selection plays in the appearance and success of R weeds.

Weed science knowledge has expanded dramatically with the appearance of herbicide resistance. Awareness of weed population diversity, intra- and inter-specific, has become a more central consideration for many researchers. Studies on s-triazine resistant and susceptible plants have revealed a subtle, more nuanced, perspective on the ecological basis of plant competitiveness (see herein). Insights into the pleiotropic consequences of the genetic changes associated with R have increased awareness of the role of R in plant communities (see herein). Elucidation of sun-air-leaf signal transduction and communication systems in the *psbA* chronomutant has provided insights into sequence of events leading to photosynthetic regulation and efficiency in R (see herein).

Crop cultivar-seed intellectual property rights have created corporate limitations on R cultivar research. This secrecy has hindered the ability of public sector science to elucidate phenotypes of R crop cultivars. R weed management would profit with insights into the ecological basis of R biotype success; as it has (herein) with atrazine R biotypes. For example, has pleiotropy occurred in glyphosate R crop cultivars?

### 4.2 How agriculture changed herbicide resistance

To understand how agriculture changed herbicide resistance it is helpful to see these phenomena from an anthropological perspective: how does human social culture, human agriculture value systems, human behavior affect weed populations? Key insights are gained by understanding the R phenotype and how human cultural activity has extended the reach of R biotypes beyond that possible without human intervention (figure 5). The concept of the 'extended phenotype' provides insight.

Richard Dawkins introduced the concept of the extended phenotype in 1999:

> **extended phenotype:** all effects of a gene upon the world; 'effect' of a gene is understood as meaning in comparison with its alleles; the concept of phenotype is extended to include functionally important consequences of gene differences, outside the bodies in which the genes sit; in practice it is convenient to limit 'extended phenotype' to cases where the effects influence the survival chances of the gene, positively or negatively.

The extended phenotype of atrazine resistant weeds are all the effects of the R gene upon the world. The concept of phenotype is extended to include functionally important consequences of gene differences *outside* the bodies in which the genes sit, reside. In practice it is convenient to limit 'extended phenotype' to cases where the effects influence the survival chances of the R gene, positively or negatively (e.g. HRC's, new R weed biotypes naturally selected in human HRC's).

> **central theorem of the extended phenotype:** an animal's behavior tends to maximize the survival of the genes "for" that behavior, whether or not those genes happen to be in the body of the particular animal performing it (Dawkins, 1999)

The central theorem of the extended phenotype is that an organism's behavior tends to maximize the survival of the genes "for" that behavior, whether or not those genes happened to be in the same body of the particular organism performing it. Human genes reside in an organism other than the body of an R plant. Human genes are involved directly with survival of R genes (e.g. farmer use of herbicides, human introduction of HRC's).

**Figure 5.** The extended phenotype of herbicide resistant weeds: herbicide discovery leading to R biotypes, then resistance mechanisms utilized in R crop cultivars, then new R weed biotypes naturally selected in resistant crops, *ad infinitum.*

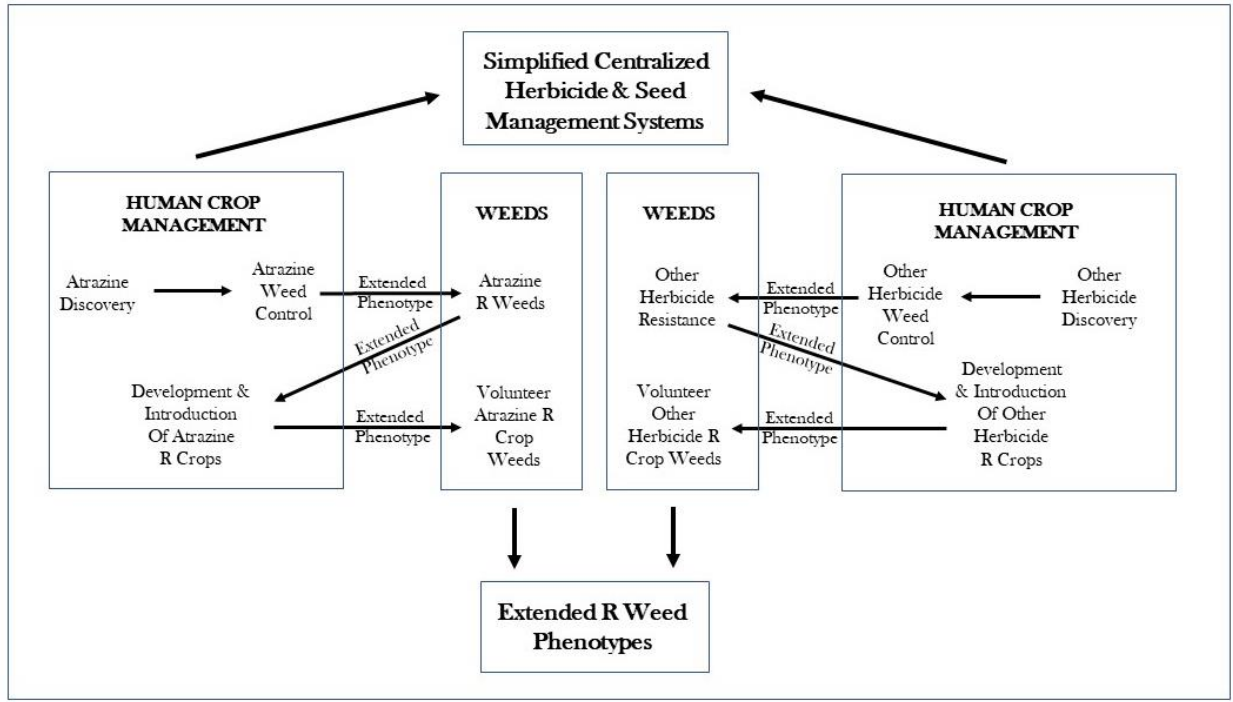